\documentclass[aps,prc,onecolumn,superscriptaddress,showpacs,showkeys]{revtex4}
\usepackage{graphicx}
\usepackage{bm}
\usepackage{epsfig}
\usepackage{amsfonts}
\usepackage{amssymb,amscd}
\usepackage{subfigure}
\usepackage{xcolor}
\newcommand{\rr}{\mbox{\boldmath $r$}}
\newcommand{\rb}{\mbox{\boldmath $b$}}

\begin{document}
%\begin{flushright}
%MS-TP-23-04
%\end{flushright}

\title{Isolated photon production in $pp$ collisions  at forward rapidities \\ and high multiplicity events}

\author{Yuri N. {\sc Lima}}
\email{limayuri.91@gmail.com}
\affiliation{Institute of Physics and Mathematics, Federal University of Pelotas, \\
  Postal Code 354,  96010-900, Pelotas, RS, Brazil}

\author{Andr\'e V. {\sc Giannini}}
\email{AndreGiannini@ufgd.edu.br}
\affiliation{Faculdade de Ci\^encias Exatas e Tecnologia, Universidade Federal da Grande Dourados (UFGD),
Caixa Postal 364, CEP 79804-970 Dourados, MS, Brazil}
\affiliation{Instituto de Ci\^encia e Tecnologia, Universidade Federal de Alfenas, 37715-400 Po\c{c}os de Caldas, MG, Brazil}

\author{Victor P. {\sc Gon\c{c}alves}}
\email{barros@ufpel.edu.br}
\affiliation{Institut f\"ur Theoretische Physik, Westf\"alische %Wilhelms-Universit\"at M\"unster,
Wilhelm-Klemm-Stra\ss e 9, D-48149 M\"unster, Germany}
%\affiliation{Institute of Modern Physics, Chinese Academy of Sciences,
%  Lanzhou 730000, China}
\affiliation{Institute of Physics and Mathematics, Federal University of Pelotas, \\
  Postal Code 354,  96010-900, Pelotas, RS, Brazil}

%----------------------------------------------------------------------
\begin{abstract}
The production of isolated photons in high multiplicity events is investigated considering the Color Glass Condensate (CGC) formalism. The associated cross section for proton - proton collisions is estimated considering three distinct solutions of the Balitsky - Kovchegov (BK) equation and predictions  for the normalized photon yield as a  function of the multiplicities of co - produced charged particles are presented. We predict  the increasing of the yield with the multiplicity, with the slope being smaller  for larger rapidities. As the isolated photon production is not affected by the fragmentation process, a future experimental investigation of this process in current high energy hadronic colliders is ideal to test the treatment of high multiplicity events using the CGC formalism, previously applied only for the production of hadronic final states.
\end{abstract}

\keywords{Photon production, Color Glass Condensate framework, High multiplicity events}
\maketitle
\date{\today}

%----------------------------------------------------------------------
\section{Introduction}

The production of isolated photons in hadronic collisions at high energies is one the cleanest probes of the strong interactions and the structure of hadrons (For recent reviews see e.g. Refs. \cite{Blau:2023bvi,David:2019wpt}). Over the last decades, this process has been largely studied, mainly motivated by the fact that, at leading order (LO), the process is dominated by the Compton scattering $q + g \rightarrow q + \gamma$, which implies that the isolated photon production is sensitive to the gluon distribution at small values of the Bjorken - $x$ variable \cite{Aurenche:1988vi, Vogelsang:1995bg, BrennerMariotto:2007yf,BrennerMariotto:2008st,Arleo:2011gc,dEnterria:2012kvo,Helenius:2014qla,Klasen:2017dsy,Goharipour:2018sip}  and to the description of the QCD dynamics at high energies \cite{
Kopeliovich:1998nw,Gelis:2002ki,Kopeliovich:2007yva,Kopeliovich:2009yw,Machado:2008nz,amir,Basso:2015pba,Ducloue:2017kkq,Benic:2016uku,Benic:2017znu,Benic:2018hvb,Goncalves:2020tvh,SampaiodosSantos:2020lte,Golec-Biernat:2020cah,Benic:2022ixp}. Another important aspect, is that  contribution from photons generated by the collinear fragmentation of final - state partons is significantly reduced, suppressing the impact of final - state interactions and making the isolated photon production an excellent probe of the wave function of incoming particles.

At high energies (or at very small $x$), the hadronic wave functions are characterized by a large number of gluons \cite{hdqcd}. Such dense system can be described by the Color Glass Condensate (CGC)  effective theory \cite{CGC}, with the evolution of the gluon distribution being given, in the mean - field approximation, by the Balitsky - Kovchegov (BK) equation \cite{BAL,KOVCHEGOV}. 
This framework implies  the limitation on the maximum phase-space parton density that can be
reached in the hadron wave function (parton saturation) and the emergence of a semihard saturation scale $Q_s(x)$, which is energy and atomic number dependent. Such scale characterizes the transition between the linear and non - linear regimes of the QCD dynamics and defines  the main aspects of the particle production in minimum bias events. Moreover, the CGC framework also predicts rare parton configurations in the hadronic wave function, which are characterized by larger values of $Q_s$ and generate high multiplicity events. As this formalism  is a promising approach for the particle production in low and high multiplicity events at the LHC, several authors \cite{Ma:2018bax,Levin:2019fvb,Kopeliovich:2019phc,Gotsman:2020ubn,Siddikov:2020lnq,Siddikov:2021cgd,Stebel:2021bbn,Salazar:2021mpv,Lima:2022mol} have applied it to  describe the event activity dependence of 
 $J/\Psi$, $D$, $K_S^0$ and $\Lambda$ yields measured in  proton - proton ($pp$) collisions, which are observed to  grow rapidly as a function of the multiplicities of co - produced charged particles \cite{ALICE:2015ikl,ALICE:2017wet,STAR:2018smh,ALICE:2020msa,ALICE:2020eji,ALICECollaboration:2020, ALICE:2021zkd}.
These studies demonstrate that the CGC framework provides a satisfactory description of the current data for low and medium multiplicities, with the predictions overshooting  the data when the multiplicity becomes very high, indicating that new ingredients in the formalism are needed  and/or new dynamical effects become relevant in these rare events. One possibility is  the modification of the hadronization process in high multiplicity events, which can suppress the yields of hadronic final states. However, its modelling is not an easy task due to the dominance of non - perturbative contributions and eventual medium effects that may be present. In addition, it is important to have a clear picture of what are the limitations of the current treatment of high multiplicity events using the CGC framework. Both aspects motivate the study of the isolated photon production in these rare events. As discussed above, the associated cross section is not affected by fragmentation processes while being sensitive to the non - linear effects in the hadronic wave function. Therefore, a future experimental analysis of this process will be ideal to disentangle the contribution of initial and final state effects in high multiplicity events and will  help us to establish the necessary improvements in the theory. In what follows, we will perform, for the first time, the analysis of the isolated photon production in high multiplicity events at $pp$ collisions using the CGC framework and present predictions for the normalized photon yield for different rapidities and distinct ranges of the photon transverse momentum. The results will be shown for  different solutions for the BK equation, making it possible to investigate different descriptions of the initial hadronic wave function. Our goal in this study is twofold. Provide the CGC predictions for the isolated photon production at high multiplicity events and motivate the experimental analysis of this process in a near future.

This paper is organized as follows. In the next Section we present a brief review of the formalism needed to describe the photon production in the Color Glass Condesante formalism. Moreover, the typical values of the projective and target momentum fractions are estimated considering different values for the rapidity and transverse momentum of the photon, which indicate the kinematical range probed in the photon production at the LHC. In Section \ref{sec:res} we compare our predictions for the photon spectra, derived considering different initial conditions for the BK equation, with the current LHC data for different rapidity ranges. The  normalized photon yield is estimated and its dependence on the multiplicity is investigated for different values of the factorization scale and rapidity as well as for distinct solutions of the BK equation. Finally, in Section \ref{sec:sum} we summarize our main results and conclusions. 

\section{Formalism}

\begin{figure}[t]
	\centering
	\includegraphics[scale=0.4]{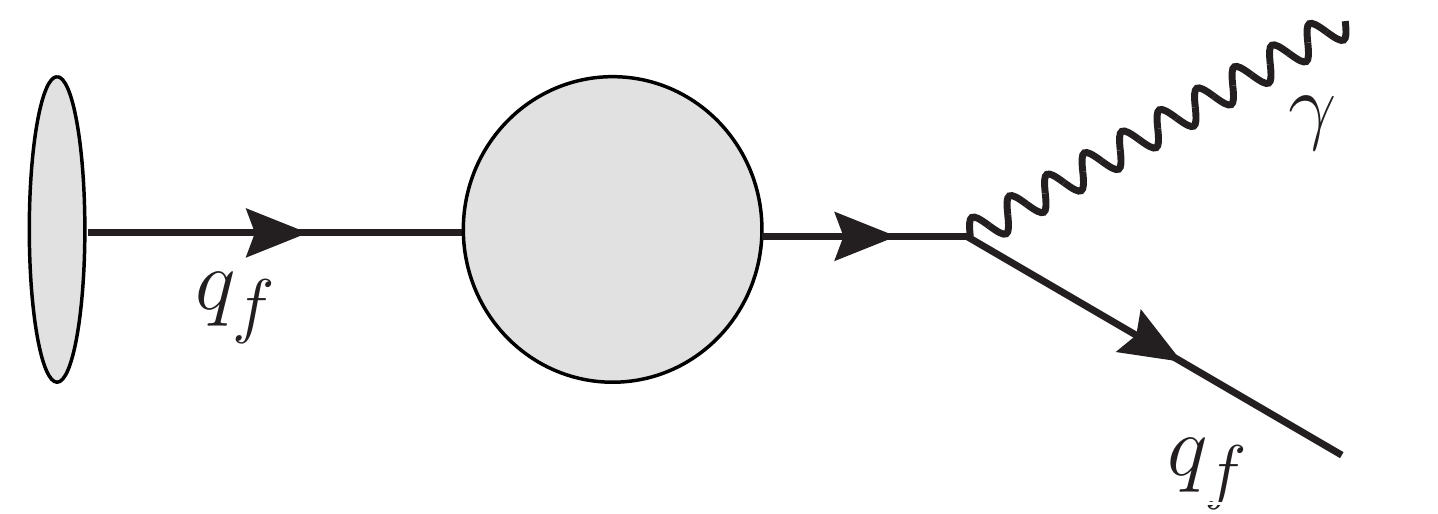}
	\hspace{2cm}
	\includegraphics[scale=0.4]{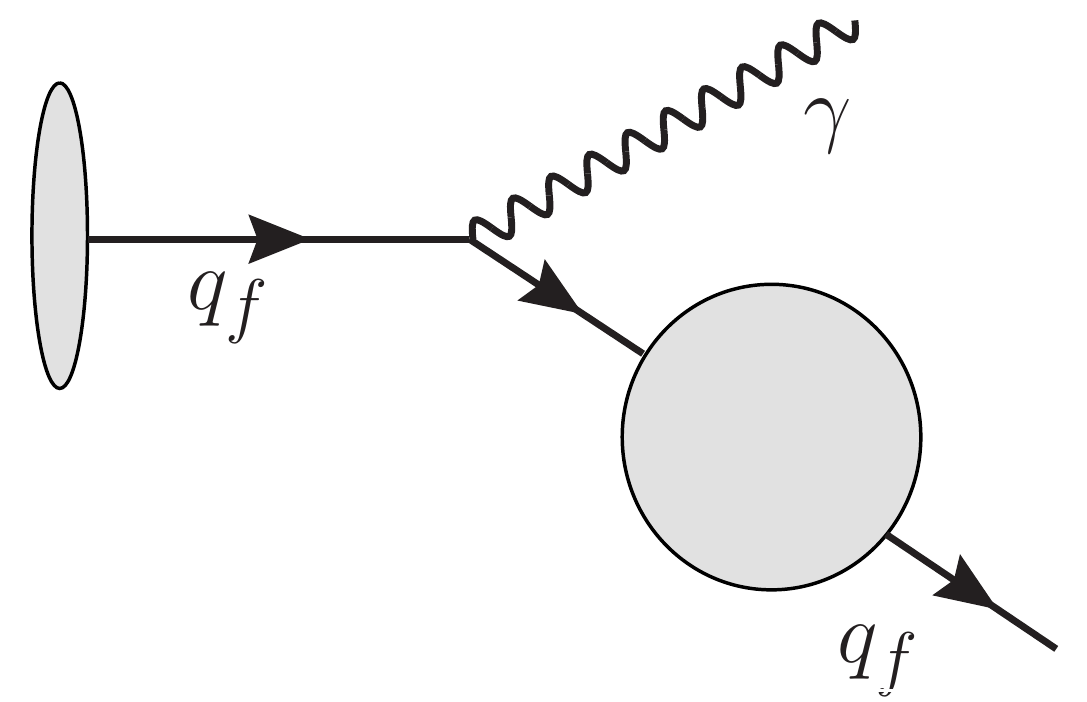}
	\caption{Process contributing for the photon bremsstrahlung off a fast projectile quark propagating through the low-$x$ color field of the target, described using the Color Glass Condensate formalism.}
	\label{Fig:diagram}
\end{figure} 

In the standard collinear formalism, the cross section for the isolated photon production at high energies is dominated by the quark - gluon ''Compton" process $qg \rightarrow q \gamma$ and, as a consequence, appears as an excellent observable to determine the gluon distribution. However, the inclusion of non - linear effects, associated e.g. to multiple scatterings, which are expected to become important at large energies and high multiplicities, is not an easy task in this formalism. An alternative is to describe this process in the Color Glass Condensate formalism, where the isolated photon production is considered as a $\gamma$  bremsstrahlung off a fast projectile quark propagating through the low-$x$ color field of the target \cite{Kopeliovich:1998nw,Gelis:2002ki}, with the photon radiation occurring either after or before the quark scatters off the target (See Fig. \ref{Fig:diagram}).  This formalism predicts that in the coordinate space the associated differential cross section can be expressed in terms of the light-cone (LC) wave function $\Psi_{q\gamma}$, describing the real 
photon radiation off the projectile quark, and the   dipole $S$ matrix, $S(x_g,\rr,\rb)$, which describes the dipole - target scattering interaction for a dipole separation $\rr$ and impact parameter $\rb$. Such quantity can be expressed in terms of the dipole - target  scattering amplitude ${\cal{N}}(x_g,\rr,\rb)$, $S = 1 - {\cal{N}}$, with the evolution in $x_g$ being obtained by solving the Balitsky - Kovchegov (BK) equation. 
One has that in the transverse momentum space and in the massless quark limit, the inclusive photon yield in a $pp$ collision is given at leading order (LO) by  \cite{Ducloue:2017kkq} 
\begin{equation}
    \frac{dN^{pp\to\gamma X}}{d^2\mathbf{k}_Tdy}=\sum_q\frac{e^2_q\alpha_{em}}{\pi(2\pi)^3}\int d^2\mathbf{l}_T\int_{x_{min}}dx_pz^2[1+(1-z)^2]\frac{q(x_p,\mu^2)}{\mathbf{k}^2_T}\frac{(\mathbf{k}_T+\mathbf{l}_T)^2}{[z\mathbf{l}_T-(1-z)\mathbf{k}_T]^2} \int d^2 \rb S(\mathbf{k}_T+\mathbf{l}_T,\rb, x_g)\,,
    \label{inclusivephotonyield}
\end{equation}
where $\mathbf{k}_T$ and $y$ are the photon transverse momentum and rapidity, respectively, $z$ represents the longitudinal momentum fraction of the quark carried by the photon, $x_{min} = k_T e^{y}/\sqrt{s}$, $q(x_p,\mu^2)$ is the collinear quark distribution function for a hard scale $\mu^2$ and the $S$ matrix in the momentum space is the Fourier transform of $S(x_g,\rr,\rb)$. 
Moreover, one has that
\begin{eqnarray}\label{eq:kinematics_photon}
    x_g=\frac{|\mathbf{k}_T|e^{-y}+|\mathbf{l}_T|e^{-y_q}}{\sqrt{s}},  \qquad y_q=\log\left(\frac{-e^{y}|\mathbf{k}_T| + x_p\sqrt{s}}{|\mathbf{l}_T|}\right), \qquad z=\frac{|\mathbf{k}_T|}{x_p\sqrt{s}}e^{y}.
\end{eqnarray}
It is important to emphasize that  over the last years, several groups
have improved the treatment of the particle production at forward rapidties in the
CGC formalism, by estimating higher-order corrections for the evolution of the forward dipole-target scattering amplitude and the associated impact factors (see, e.g.,
Refs. \cite{Chirilli:2011km,Balitsky:2012bs,Altinoluk:2014eka,Beuf:2016wdz,Iancu:2016vyg,Lappi:2016fmu,Boussarie:2016ogo,Beuf:2017bpd,Boussarie:2016bkq,Caucal:2021ent,Liu:2022ijp,Taels:2022tza,Bergabo:2022tcu,Fucilla:2022wcg,Bergabo:2022zhe,Bergabo:2023wed,Caucal:2023nci,Altinoluk:2023hfz,Fucilla:2023mkl,Taels:2023czt}). In particular, the forward production of a Drell-Yan pair and a jet was recently estimated at next-to-leading order in Ref. \cite{Taels:2023czt}. In principle, such a result can be used to derive the NLO corrections for the $q \rightarrow q + \gamma$ impact factor needed to estimate the isolated photon production.  
Surely, a detailed comparison between our results and those derived at the NLO is an
important next step, which we plan to perform in a future study. However, in principle, we do not expect a
large modification of our predictions for high multiplicities if the NLO corrections
are taken into account, since in our analysis we are interested in normalized yields, $dN_{\gamma}/dy  / \langle dN_{\gamma}/dy \rangle$, where  $\langle dN_{\gamma}/dy \rangle$ is the yield for minimum bias events.

In what follows, in order to reduce  the fragmentation component for the photon production, we will enforce an isolation cut by multiplying the integrand of Eq. (\ref{inclusivephotonyield}) by $\theta[\sqrt{(y - y_q)^2 + \Delta \phi} - R]$, where $\Delta \phi$ is the azimuthal angle between the scattering quark and the photon and $R$ is a chosen isolation cone radius. 
As in Ref. \cite{Ducloue:2017kkq}, we will use the leading order CTEQ6 parton distribution functions \cite{cteq} to describe the quark content of the projectile and  assume that the impact parameter in the $S$ matrix  can be factorized, such that 
\begin{eqnarray}
\int d^2 \rb \, {\cal{N}}(x_g,\rr,\rb) = \frac{\sigma_0}{2} \, N(x_g,\rr) \,\,,
\end{eqnarray}
where $\sigma_0$ is a constant determined by fitting the HERA data for the reduced cross section and $N(x_g,\rr)$ is obtained by solving the running coupling BK equation \cite{Albacete:2007yr,Albacete:2009fh,Albacete:2010sy}.  In our analysis we will consider three different solutions for the rcBK equation obtained assuming different parametrizations for the initial condition $N(x_g = 0.01,\rr)$ and that provide similar values of $\chi/d.o.f$ for the HERA data \cite{Albacete:2012xq}. In particular, we will consider the solution derived assuming the GBW parameterization, $N(x_g = 0.01,\rr) = 1 - 
\exp[-\frac{\rr^2 Q_{s,0}^2}{4}]$ with  $Q_{s,0}^2 = 0.24 $ GeV$^2$,      and  two solutions obtained assuming a MV - type initial condition $N(x_g = 0.01,\rr) = 1 - \exp[-\frac{(\rr^2 Q_{s,0}^2)^{\gamma}}{4} \ln (\frac{1}{\Lambda_{QCD} r}+ e)]$, characterized by ($Q_{s,0}^2 = 0.157$ GeV$^2$, $\gamma = 1.101$) and ($Q_{s,0}^2 = 0.1597$ GeV$^2$, $\gamma = 1.118$). 
In what follows, the predictions originating from these distinct input will be denoted, respectively, by g1 (GBW), g1.101 (MV) and   g1.118 (MV). For completeness, in the analysis of the photon spectra, we will also present the predictions derived assuming $\gamma = 1.0$ in the MV initial condition, with the prediction being denoted g1 (MV) hereafter.

\begin{figure}[t]
	\centering
	\includegraphics[scale=0.7]{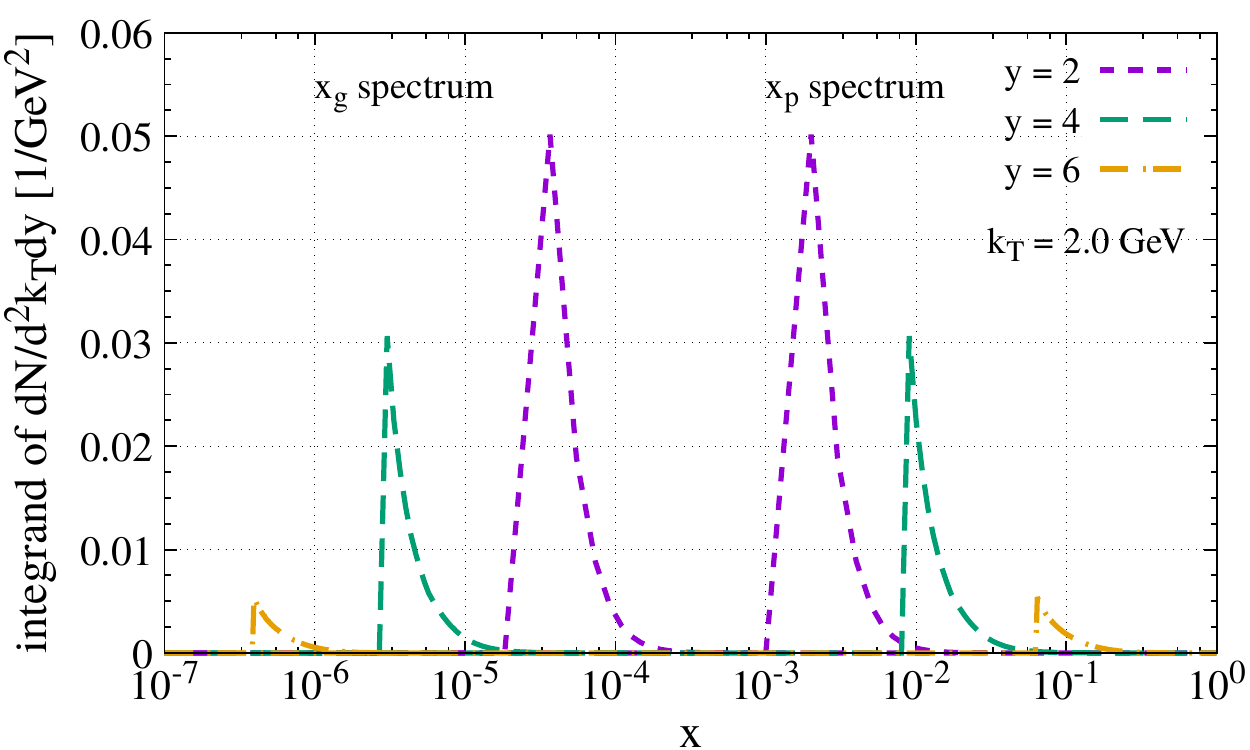}
	\includegraphics[scale=0.7]{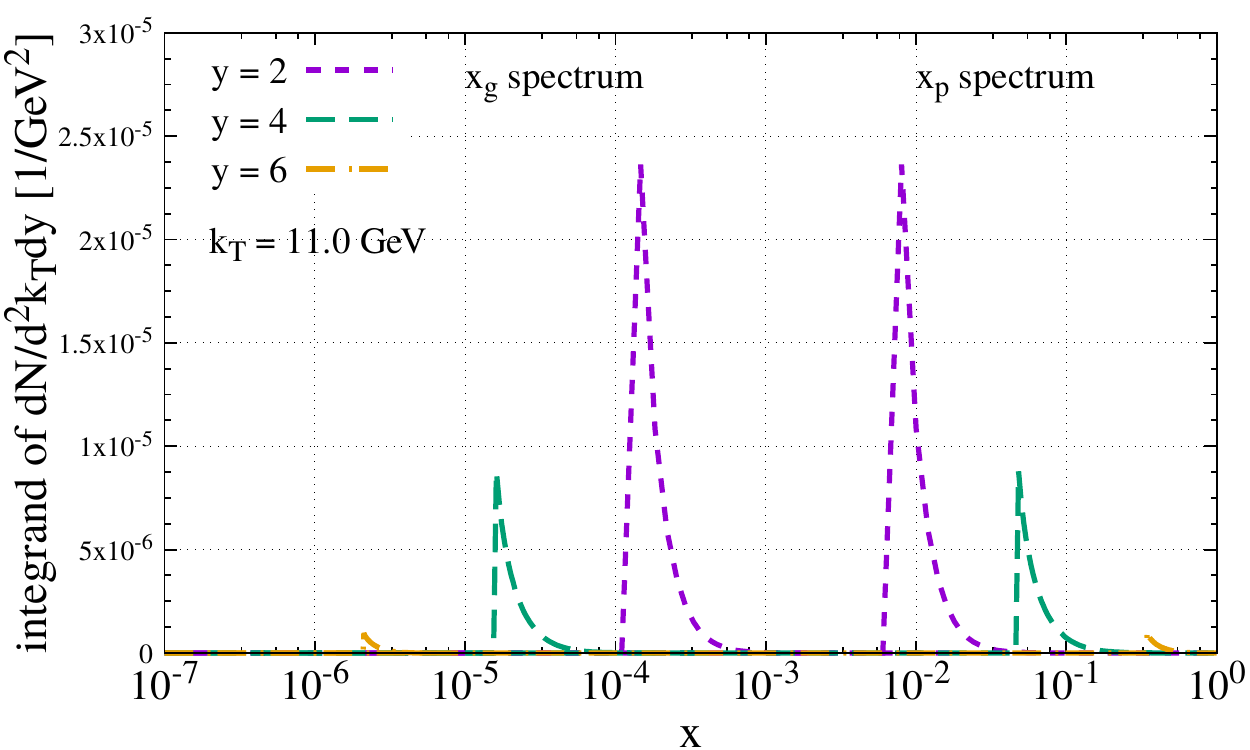}
	\caption{The momentum fractions of projectile ($x_p$) and target ($x_g$) partons contributing to the
cross section from Eq. (\ref{inclusivephotonyield}) for different values of the rapidity and transverse momentum of the photon. Results derived for $pp$ collisions at $\sqrt{s} = 13$ TeV.}
	\label{Fig:integrand}
\end{figure}

The basic assumption in the CGC formalism for the isolated photon production is that the projectile evolves according to the linear DGLAP dynamics and the target is treated using the CGC
formalism. In other words, such an approach is expected to be valid when the  cross section is  dominated by collisions of projectile
partons with large light cone momentum fractions with target partons carrying a very small momentum fraction.
In Fig. \ref{Fig:integrand} we present an estimate of the typical momentum fractions of projectile ($x_p$) and target ($x_g$) partons contributing to the
cross section from Eq. (\ref{inclusivephotonyield}), derived considering $pp$ collisions at $\sqrt{s} = 13$ TeV. We assume different values for the rapidity and two values for the photon transverse momentum: $k_T = 2.0$ GeV (left panel) and  $k_T = 11.0$ GeV (right panel).
The results clearly demonstrate that for large rapidities and high transverse momentum, the assumption present in the CGC formalism is fully satisfied. For lower values of rapidity and transverse momentum, 
one has that smaller values of  projectile momentum fractions starts to contributes and non-linear effects can become non - negligible. In order to reduce the impact of these corrections,  we will consider $k_T \ge 4.0$ GeV in the analysis of photon production at high multiplicities. It is important to emphasize that for central rapidities one has $x_p \approx x_g$, which implies that the application of the CGC formalism for the description of the photon production is theoretically questionable. Therefore, in what follows, we will restrict our analysis for forward rapidities.
%Although aware of this aspect, in our analysis  we will compare the CGC predictions with the current experimental LHC data for central rapidities, in order to analyze whether this approach provides  a good approximation for the description of the process in this kinematical region.

\begin{figure}[t]
	\centering
	\includegraphics[scale=0.7]{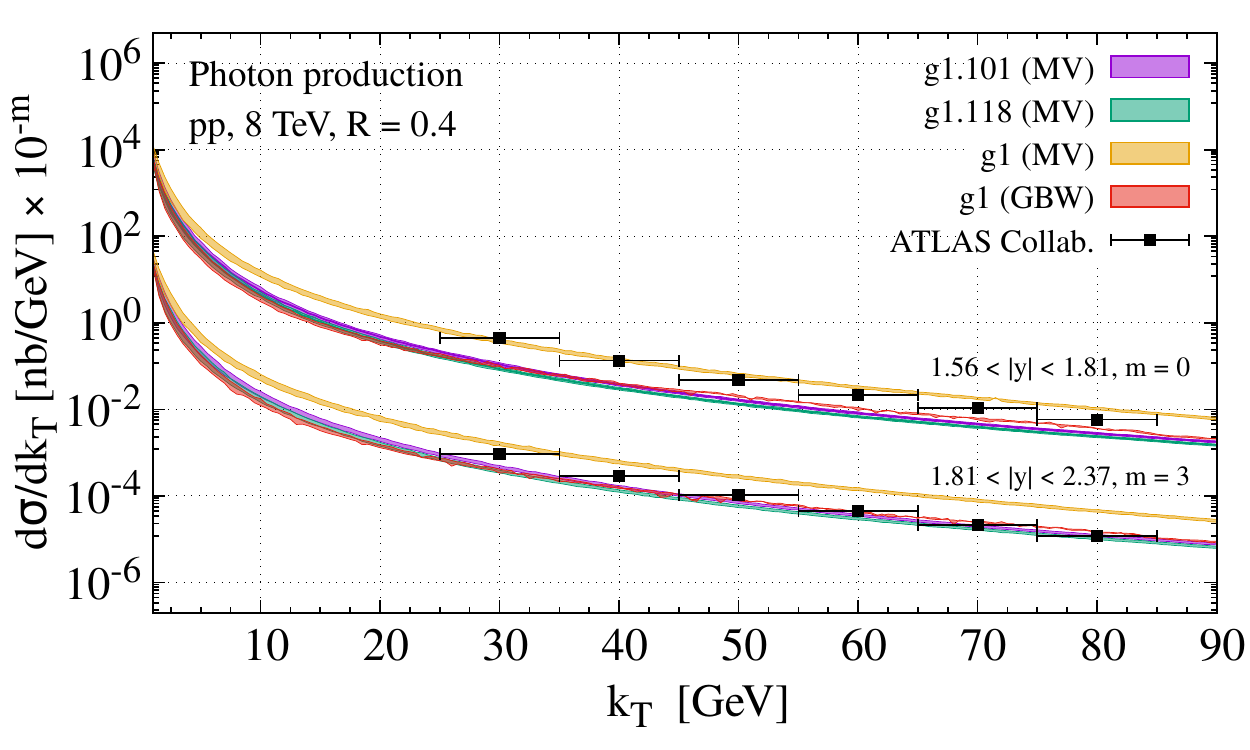}
	\includegraphics[scale=0.7]{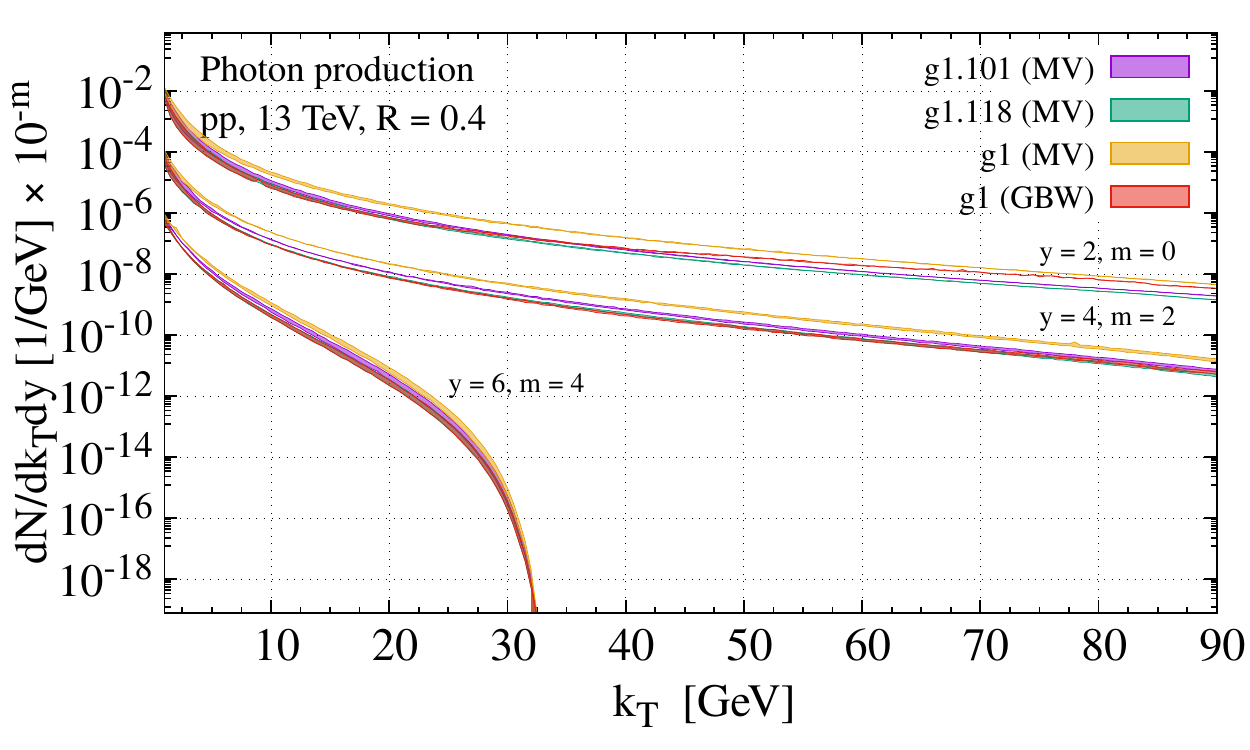}
	\caption{Predictions of the CGC formalism for the isolated photon production in $pp$ collisions for different center - of - mass energies derived considering distinct solutions of the BK equation.   {\bf Left panel:} Comparison with the ATLAS data for the photon spectra  $d\sigma/dk_T$ integrated over different rapidity ranges.
{\bf Right panel:} Predictions for the invariant yield for fixed values of rapidity. The predictions and data for the distinct rapidities presented in the lower panels have been shifted by a multiplicative factor $10^{-m}$, with the value of $m$ shown on the plots.}
	\label{Fig:spectra}
\end{figure}

\section{Results}
\label{sec:res}
Initially, let's present our predictions for the photon transverse momentum spectra for different center - of - mass energies and distinct values of rapidity. In our calculations we will make use of the distinct solutions of the BK equation for the dipole - proton scattering amplitude discussed in the previous section and  show  results derived by varying the factorization scale in the range $0.25 Q^2 \le \mu^2 \le 4 Q^2$, with $Q^2 = \mbox{max}\{\mathbf{l}_T^2,\mathbf{k}_T^2\}$ \cite{Ducloue:2017kkq}. 
In Fig. \ref{Fig:spectra} (left  panel) we present our predictions for the differential distribution, $d\sigma/dk_T$, integrated over distinct ranges of rapidity, and the current ATLAS data \cite{ATLAS:2016} for the isolated photon production. Moreover,  we have assumed $R = 0.4$. One has that the g1(MV) prediction implies a larger normalization in comparison with the other solutions, overestimating the current experimental data. In contrast, the predictions derived using the other solutions satisfactorily describe the data, in particular
 for larger rapidities. It is important to emphasize that an overall multiplicative $K$-factor, usually present in predictions originating from calculations based on the collinear formalism, is not present in our predictions. 
%In Fig. \ref{Fig:spectra} (lower left panel) we present a comparison between our predictions for the photon spectra $d\sigma/dk_T$, integrated over distinct ranges of rapidity, and the current ATLAS data \cite{ATLAS:2016} for the isolated photon production. As before, the data are overestimated by the g1(MV) prediction and satisfactorily described by the other solutions. In particular, these solutions describe quite well the data for larger rapidities. 
In Fig. \ref{Fig:spectra} (right panel), we present  our results for the transverse momentum dependence of the invariant photon yield for $pp$ collisions at $\sqrt{s} = 13 $ TeV  assuming distinct values for the photon rapidity. One has that the predictions differ mainly at large values of $k_T$. It is important to emphasize that for $y = 6$ one has a strong reduction of the distribution for large tranverse momentum due to the smaller phase space available in comparison to smaller values of $y$.  The comparison of these predictions with future experimental data will be useful to check the validity of the formalism, especially at forward rapidities where we expect a larger impact of the non - linear effects on the QCD dynamics. Considering the above results, in what follows we will not present the predictions derived assuming the g1(MV) initial condition.

\begin{figure}[!th]
	%\centering
	\includegraphics[scale=0.75]{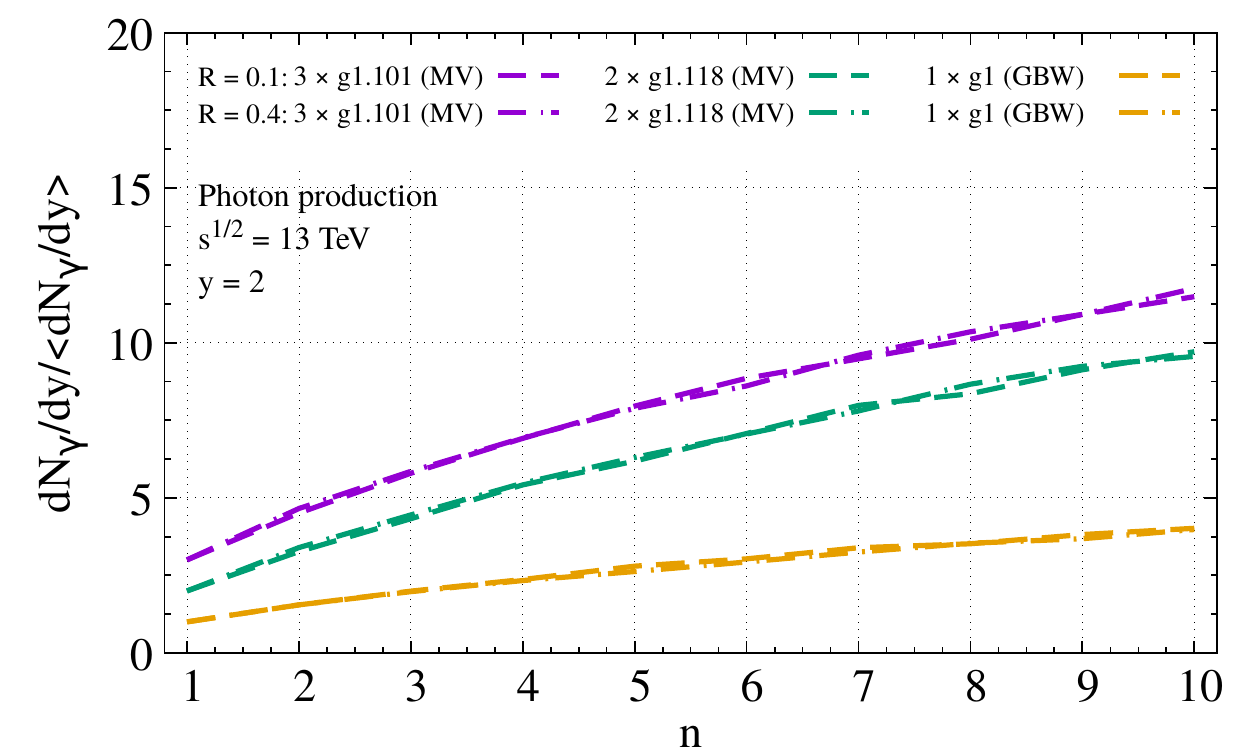}
	\includegraphics[scale=0.73]{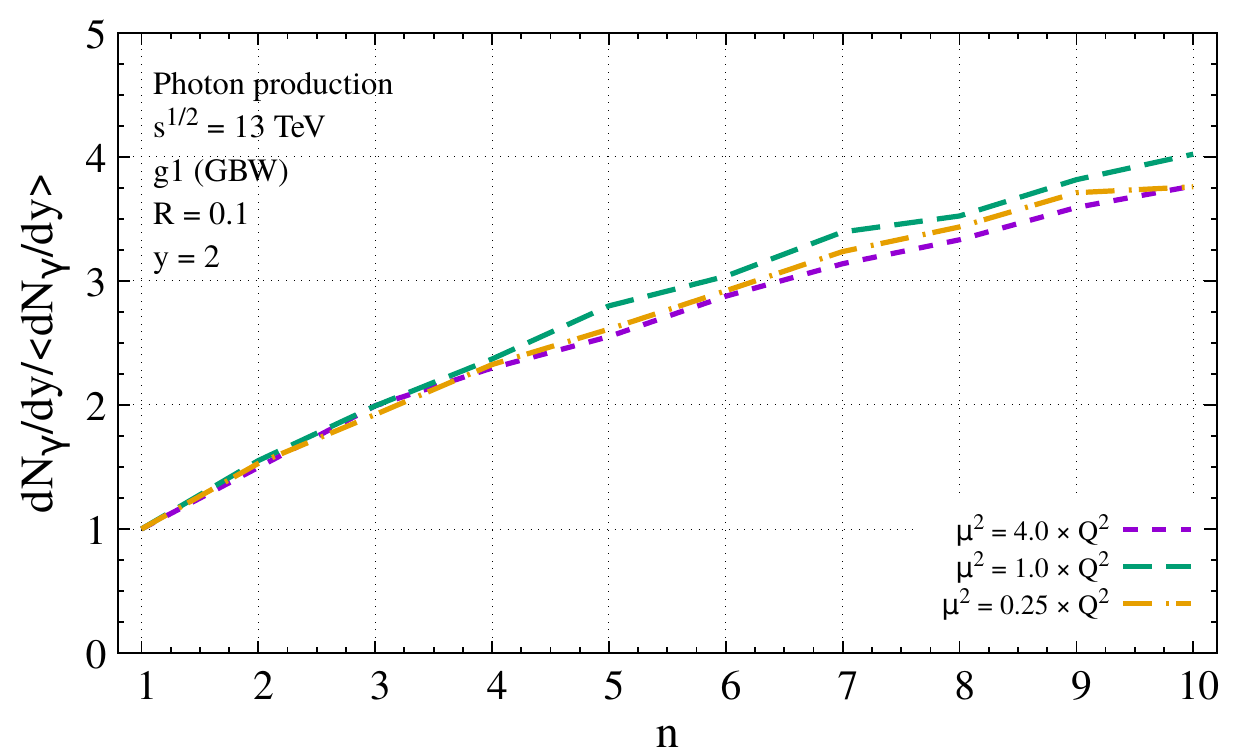}
	\includegraphics[scale=0.75]{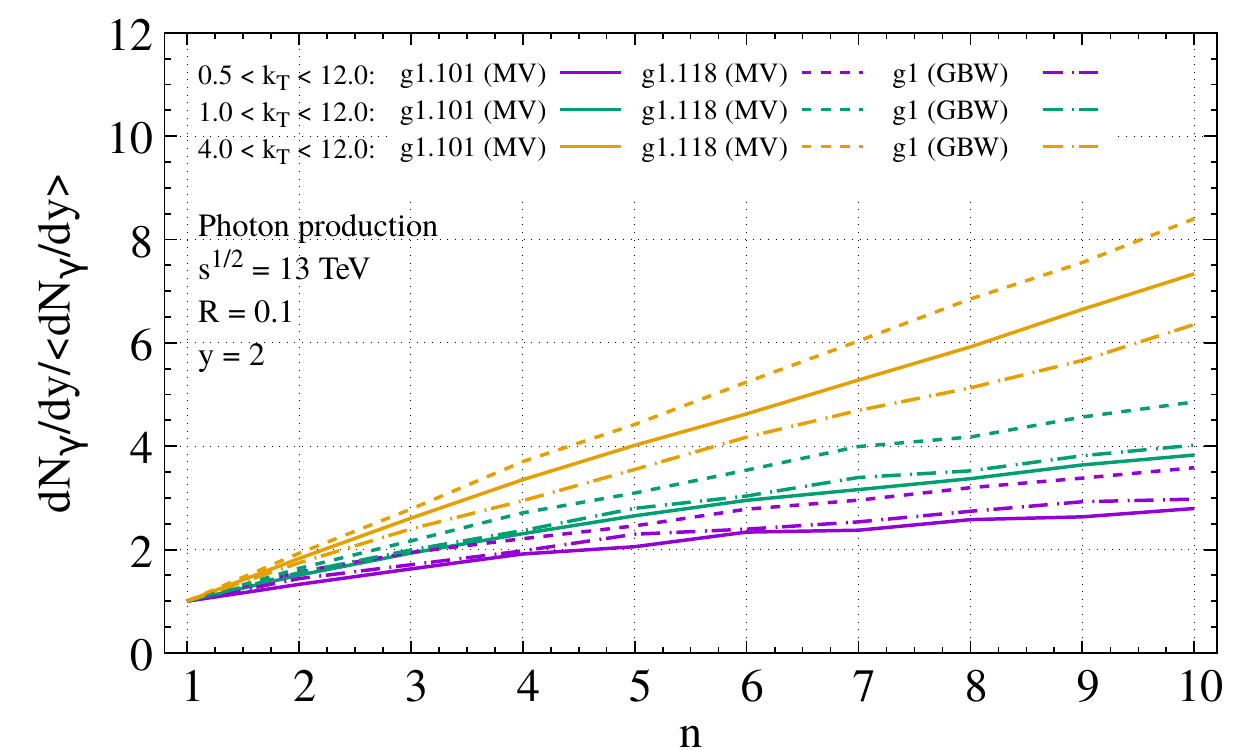}
	\caption{Dependence of the normalized photon yield on the scale factor $n$, which defines the value of the saturation scale in the initial condition of the BK equation, for different values of $R$ (top panel), of the factorization scale (middle panel) and for distinct integration ranges of the photon transverse momentum (bottom panel).}
	\label{Fig:dnphoton}
\end{figure}

\begin{figure}[t]
	\centering
		\includegraphics[scale=0.68]{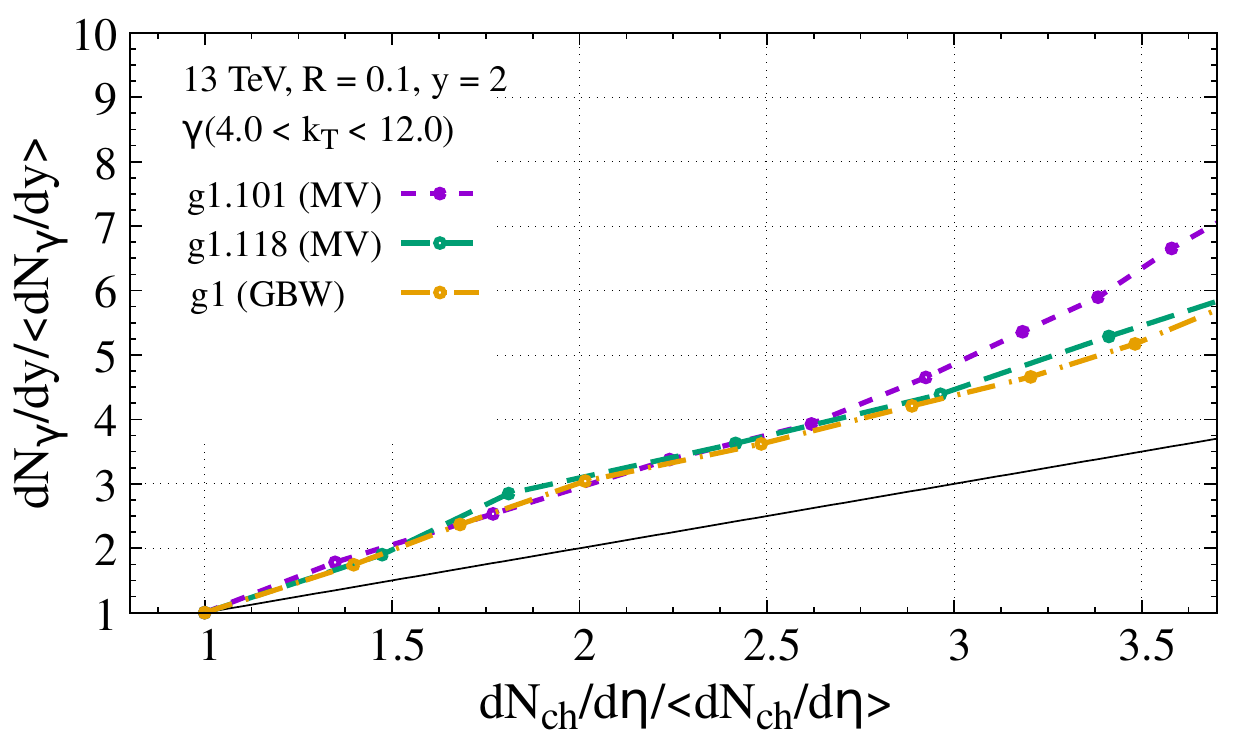}
		\includegraphics[scale=0.68]{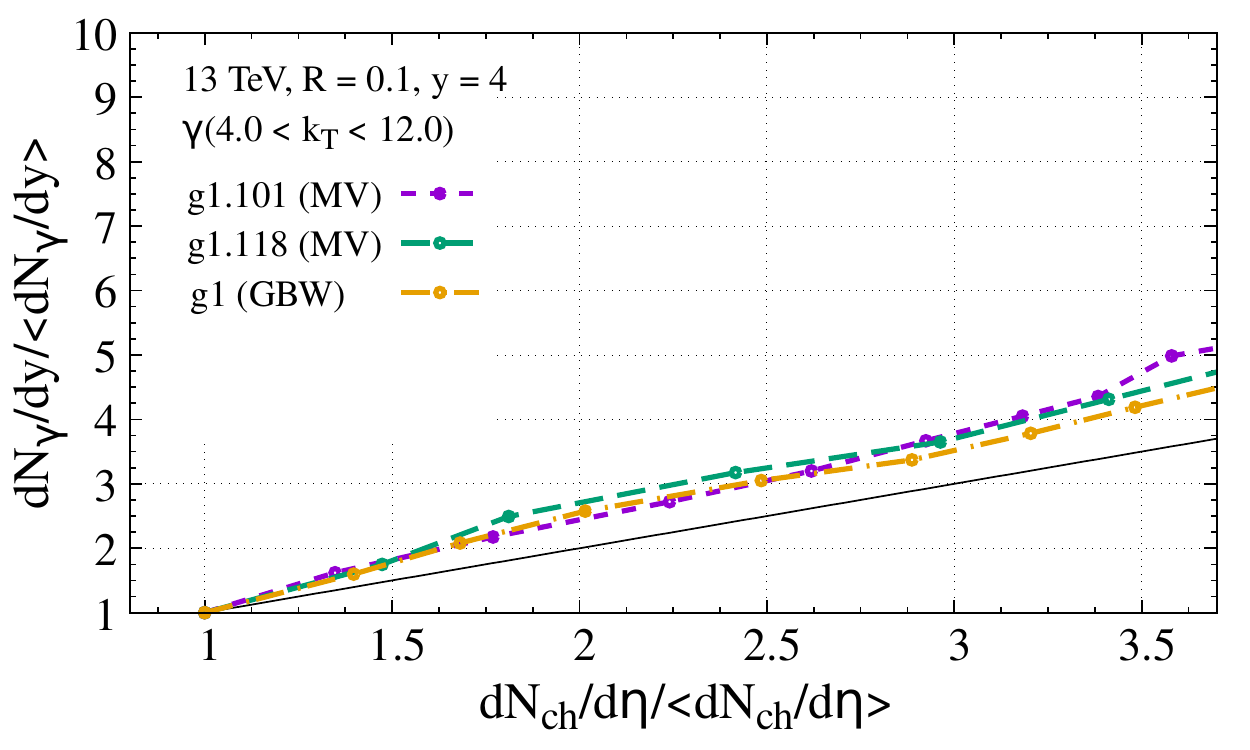}
		\includegraphics[scale=0.68]{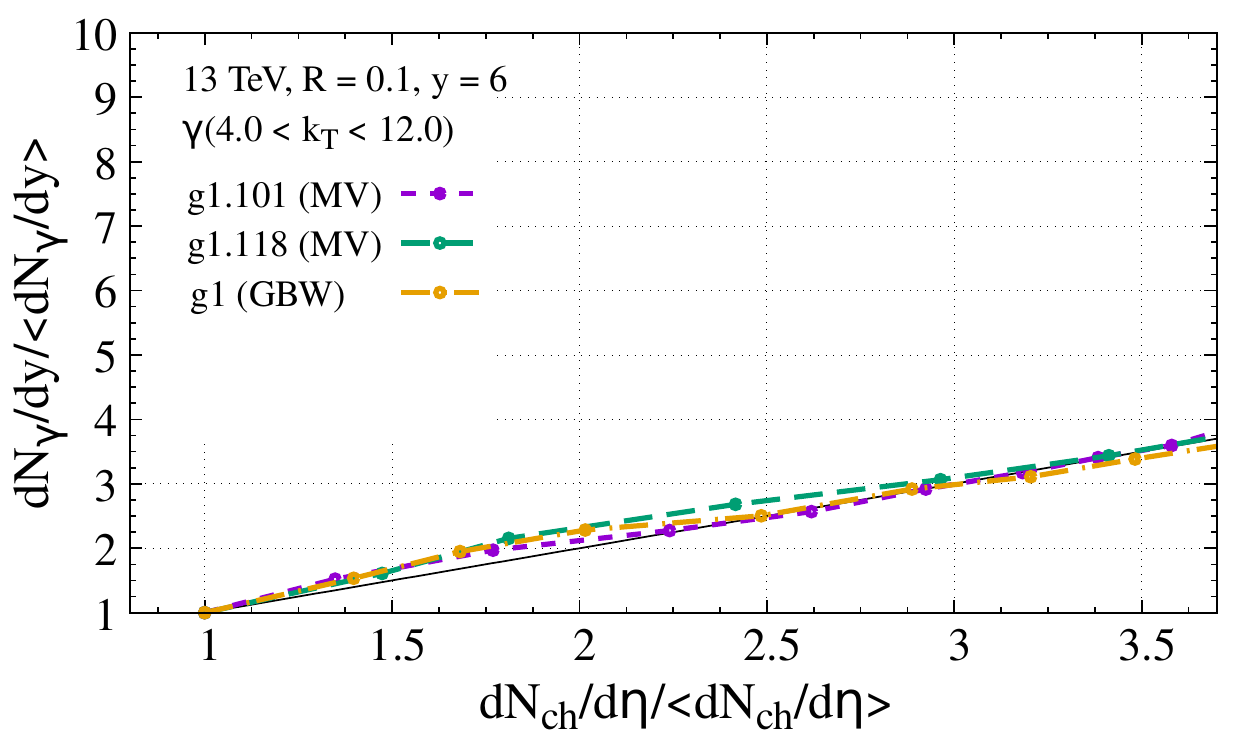}
	\caption{Correlation between the normalized isolated photon and charged particles yields in $pp$ collisions at $\sqrt{s}=13$ TeV, for three distinct solutions of the BK equation and different values of the photon rapidity.}
	\label{Fig:correlation}
\end{figure}

In what follows we will focus on the isolated photon production in high multiplicity events at the LHC. Our goal is to estimate the ratio $dN_{\gamma}/dy  / \langle dN_{\gamma}/dy \rangle$ as a function of the multiplicity, where 
$dN_{\gamma}/dy$ is the rapidity distribution for a given multiplicity, obtained from Eq. (\ref{inclusivephotonyield}) by integrating it over a given range of $k_T$, and  $\langle dN_{\gamma}/dy \rangle$ is its minimum bias value. 
As in previous studies \cite{Ma:2018bax,Levin:2019fvb,Kopeliovich:2019phc,Gotsman:2020ubn,Siddikov:2020lnq,Siddikov:2021cgd,Stebel:2021bbn,Salazar:2021mpv,Lima:2022mol}, we will assume that the particle production mechanism is the same for low and high - multiplicity events, with the main difference being the saturation scale present in these two classes of events. Such a procedure may be justified as follows: naturally, each scattering event probes a different color charge configuration of the colliding system, which in the CGC approach is characterized by a distinct saturation scale. Evoking the general expectation, based on the local parton - hadron duality,  that the final multiplicity of a particular event correlates to the initial partonic configuration of such systems leads to the notion that highly occupied states have larger saturation scales than those of typical (minimum bias) configurations. In particular, the results derived in Ref.~\cite{Lappi:2011gu} indicated that the high multiplicity configurations can be approximated by increasing the value of $Q_s$ as follows \begin{eqnarray}
Q_{s}^2(n) = n \cdot Q_{s}^2 \,\,,
\end{eqnarray}
where $n$ characterizes the multiplicity. 
Such assumption implies then  that 
the high multiplicity configurations can be approximated by increasing the value of $Q_{s,0}^2$ in the initial condition of the BK equation [$Q_{s,0}^2(n) = n \cdot Q_{s,0}^2$] and that Eq. (\ref{inclusivephotonyield}) can be assumed to be valid for both classes of multiplicities.

Our results for the dependence on $n$ of the normalized yield are presented in Fig. \ref{Fig:dnphoton} considering the production of an isolated photon with $y = 2$ in $pp$ collisions at $\sqrt{s} = 13$ TeV and assuming distinct solutions of the BK equation. In the top and middle panels, we present results obtained by integrating the transverse momentum of the photon in the range $1.0 \le k_T \le 12$ GeV. The dependence on these values is analyzed in the bottom panel.   Initially, in the upper panel, we analyze the dependence of our predictions on the value of the isolation cone radius $R$. It is important to note that the predictions derived using the distinct solutions of the BK equation have been rescaled by different constant factors to improve visualization. One has that the predictions are almost identical for the two values of $R$ considered, which is expected since we are estimating the normalized yield. As a consequence, we will limit ourselves to show only predictions derived assuming $R = 0.1$. The dependence on the factorization scale $\mu^2$ is presented in the middle panel for the g1(GBW) solution. One has that the high $n$ behavior is sensitive to the choice for the hard scale, with larger values of $\mu^2$ implying a reduction of the normalized photon yield. A similar behavior is verified for the other solutions of the BK equation. 
Finally, in the lower panel, we present the results derived by integrating over distinct $k_T$ ranges. One has that the predictions are sensitive to the $k_T$ range and BK solution considered in the calculation. In particular, the increasing of the normalized photon yield with $n$ is steeper when the minimum value of $k_T$ ($k_T^{\min}$)  is larger. A similar behavior already been observed for other final states \cite{Ma:2018bax,Levin:2019fvb,Kopeliovich:2019phc,Gotsman:2020ubn,Siddikov:2020lnq,Siddikov:2021cgd,Stebel:2021bbn,Salazar:2021mpv,Lima:2022mol}. Moreover, the predictions also become more dependent on the BK solution used in the calculation for a larger $k_T^{\min}$. Such behaviour is directly associated to the fact that the main difference between the BK solutions occurs for larger values of the transverse momentum and this difference is amplified when larger values of initial saturation scale are assumed.

In recent years, the dependence on the multiplicity of a process has been analyzed by studying the correlation between the normalized yield for the specific final state and for charged hadrons. The latter is described by  $dN_{ch}/d\eta/ \langle dN_{ch}/d\eta \rangle$ and is predicted to be proportional to $n$ when the non - linear effects are taken into account (See e.g.  Ref. \cite{Lappi:2011gu}). In order to study this correlation for the isolated photon case,   we will estimate the normalized yield for charged particles  as in Ref. \cite{Lima:2022mol},   taking into account the contribution of charged pions, baryons and strange mesons and assuming the distinct solutions for the BK equation. As in Ref. \cite{ALICE:2021zkd}, the charged hadron yield will be estimated in all cases integrating over the transverse momentum (with $p_T^{min}$ = 0.1 GeV) and assuming that the particles are produced  at central rapidities.  One has verified that the corresponding predictions describe the current data for the inclusive hadron production at central rapidities in $pp$ collisions at the LHC. Predictions for the correlation are presented in Fig. \ref{Fig:correlation} for different values of the photon rapidity $y$ and distinct solutions of the BK equation. The results have been obtained by integrating the photon momentum in the range  $4.0 \le k_T \le 12.0$ GeV, the same range considered by the ALICE Collaboration in its measurement of the $K_S^0$ meson \cite{ALICECollaboration:2020}. Predictions for other $k_T$ ranges can be provided upon request. The solid line in Fig. \ref{Fig:correlation} indicates the expected result for a linear correlation between the yields.
One has that the increasing of the isolated photon yield with the multiplicity is strongly dependent on the rapidity, becoming weaker as $y$ increases. In particular, for very forward rapidities, we predict an almost linear dependence of the multiplicity, with the results obtained by employing different solutions of the BK equation being similar. Such a result is expected, since for large values of $y$ one has large values of the saturation scale, implying that the isolated photon production at forward rapidities in the transverse momentum  $4.0 \le k_T \le 12.0$ GeV will be impacted by the non - linear effects in the QCD dynamics as the charged hadron yields for midrapidities and $p_T \ge 0.1$ GeV.
Moreover, our results indicate that the predictions for large multiplicities and smaller rapidities are sensitive to the description of the QCD dynamics, with the g1.101 (MV) solution predicting a larger enhancement. A future comparison of these predictions with the experimental data will be an useful test of the CGC formalism as well of the main assumptions present in the treatment of the high multiplicity events.

%This expectation can be understood looking at the kinematics of the photon production (Eq. (\ref{eq:kinematics_photon})), where $x_g$ can be rewritten as:
%\begin{eqnarray}
%	x_g=\frac{|\mathbf{k}_T|e^{-y}}{\sqrt{s}} + \frac{|\mathbf{l}_T|^2}{(x_p\sqrt{s} - |\mathbf{k}_T|e^{y})s} = \frac{|%\mathbf{k}_T|e^{-y}}{\sqrt{s}} + \frac{|\mathbf{l}_T|^2}{(x_p - x_{min})s}\,. 
%\end{eqnarray}
%When $x_p \gg x_{min}$, the second term in the equation above is suppressed and $x_g$ is fixed at a constant value; such a value depends on the kinematic region probed. For lower energies (and/or central rapidities), $x_g$ receives two non-negligible contributions, increasing the minimum value of $x_g$ probed and, therefore, preventing that the saturation region is accessed. On the other hand, at high energies and in the forward region (region characterized by large saturation scale values in the target, so dominated by the small-$x$ evolution), the second term is suppressed and the kinematics of the photon production tends to the one appearing in the charged particle production. 

\begin{figure}[t]
	\centering
	\includegraphics[scale=0.9]{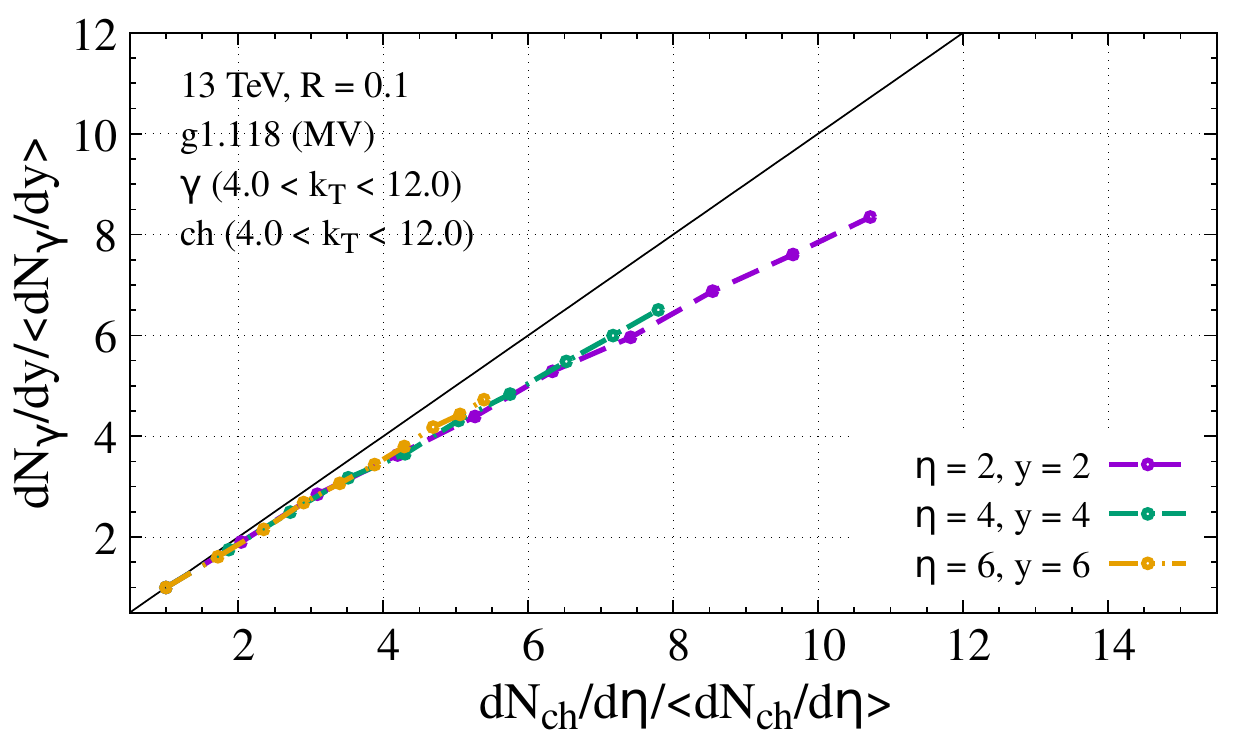}
	\caption{Correlation between the normalized isolated photon and charged particles yields in $pp$ collisions at $\sqrt{s}=13$ TeV, derived that the distinct yields are integrated over the same transverse momentum range and are estimated at the same rapidity.}
	\label{Fig:difraps}
\end{figure}

%\begin{figure}[t]
%	\centering%
%	\subfigure[]{
%		\includegraphics[scale=0.68]{Correlation_photon-charged_13TeV_5-kT-20_y0.pdf}}
%	\centering
%	\subfigure[]{
%		\includegraphics[scale=0.68]{Correlation_photon-charged_13TeV_10-kT-20_y0.pdf}}
%	\caption{Correlation between the photon and charged particles yields in $pp$ collisions at $\sqrt{s}=13$ TeV, derived considering three distinct solutions of the BK equation, named g1.101 (MV), g1.118 (MV), and g1 (GBW). For $y = 0$ and considering different ranges of the transverse momentum: (a) $5.0<k_T<20.0$ and (b) $10.0<k_T<20.0$.}
%	\label{correlations2}
%\end{figure}

%\begin{figure}[t]
%	\centering
%	\includegraphics[scale=0.9]{dNdkTdy-vs-kT_13TeV.pdf}
%	\caption{The photon transverse-momentum spectra in $pp$ collisions at $\sqrt{s}=13$TeV of the LHC experiments using  three distinct UGD models: g1.101 (MV), g1.118 (MV), and g1 (GBW). For different rapidity values ($y = 0, 2, 4, 6$). The uncertainty band considers a check depending on the saturation scale, with the upper limit being $\mu^2=4Q^2$ and the lower limit being $\mu^2=0.25Q^2$. Each set of measurements and predictions in a given rapidity bin is shifted by a multiplicative factor $10^{-m}$, wich the value of $m$ shown on the plots.}
%	\label{spectra2}
%\end{figure}

%----------------------------------------------------------------------
%\section{Results and discussion}
%\label{section:results}

In the above paragraphs we have presented our results for the correlation between the normalized isolated photon and charged particle yields for different photon rapidities derived considering that charged particles are ever produced at central rapidities and the associated yields are obtained by integrating over different transverse momentum ranges, as assumed in the analyzes performed by the ALICE Collaboration using the Run 2 data. However, for the run 3, a higher statistics is also expected for larger rapidities, which will allow to study the correlation between the yields for identical rapidities and transverse momentum ranges. Considering this perspective, in Fig. \ref{Fig:difraps} we present our predictions for the correlation derived assuming that the photon and the charged particles are produced at the same rapidity and its yields are integrated over the same transverse momentum range  ($4.0$ GeV  $\le k_T \le 12.0$ GeV). One has that for this case, the predictions are almost insensitive to the rapidity, which is expected since the isolated photon and charged hadron production have a similar dependence on the non-linear effects when the same kinematical region is probed. In addition, we predict that the isolated photon yield will be suppressed in comparison to the charged hadron one, which can be associated to the fact that the charged hadron yield has a faster enhancement with the increasing of the multiplicity. The comparison of this prediction with future experimental data provides an important test of the CGC formalism.

\section{Summary}
\label{sec:sum}

In this paper we have investigated the isolated photon production in high multiplicity $pp$ collisions at the LHC considering the CGC formalism, which provides an unified approach for the treatment of low and high multiplicity events as well for the description of the hadron and photon production at high energies. Our study has been motivated by the fact that the isolated photon yield is not expected to be affected by final state interactions and hadronization effects, which can modify the predictions for the hadron production in rare events. Therefore, a future comparison between the predictions presented in this paper with experimental data will be a clean probe of the CGC formalism and an important test of the assumptions assumed in the modelling of the high multiplicity events. We strongly motivate the experimental analysis of the isolated photon production  in $pp$ collisions at distinct multiplicities and different rapidities in the forthcoming years.

\section*{Acknowledgements}
V.P.G. thanks A. Andronic for fruitful discussions about the measurement of high multiplicity events at the LHC.
This work was partially supported by INCT-FNA (Process No. 464898/2014-5).
V.P.G. was partially supported by CNPq, CAPES and FAPERGS. Y.N.L. was partially financed by CAPES (process 001). 
A.V.G. has been partially supported by CNPq. The authors acknowledge the National Laboratory for Scientific Computing (LNCC/MCTI, Brazil), through the ambassador program (UFGD), subproject FCNAE for providing HPC resources of the SDumont supercomputer, which have contributed to the research results reported within this study. URL: \url{http://sdumont.lncc.br}

%----------------------------------------------------------------------

%----------------------------------------------------------------------

\begin{thebibliography}{99}

\bibitem{Blau:2023bvi}
D.~Blau and D.~Peresunko,
%``Direct Photon Production in Heavy-Ion Collisions: Theory and Experiment,''
Particles \textbf{6}, no.1, 173-187 (2023)

\bibitem{David:2019wpt}
G.~David,
%``Direct real photons in relativistic heavy ion collisions,''
Rept. Prog. Phys. \textbf{83}, no.4, 046301 (2020)



\bibitem{Aurenche:1988vi}
P.~Aurenche, R.~Baier, M.~Fontannaz, J.~F.~Owens and M.~Werlen,
%``The Gluon Contents of the Nucleon Probed with Real and Virtual Photons,''
Phys. Rev. D \textbf{39}, 3275 (1989)


\bibitem{Vogelsang:1995bg}
W.~Vogelsang and A.~Vogt,
%``Constraints on the proton ' $s$ gluon distribution from prompt photon production,''
Nucl. Phys. B \textbf{453}, 334-354 (1995)


\bibitem{BrennerMariotto:2007yf}
C.~Brenner Mariotto and V.~P.~Goncalves,
%``Enhancement of prompt photons in ultrarelativistic proton-proton collisions from nonlinear gluon evolution at small-x,''
Phys. Rev. C \textbf{75}, 068202 (2007)

\bibitem{BrennerMariotto:2008st}
C.~Brenner Mariotto and V.~P.~Goncalves,
%``Nuclear shadowing and prompt photons in hadronic collisions at ultrarelativistic energies,''
Phys. Rev. C \textbf{78}, 037901 (2008)


\bibitem{Arleo:2011gc}
F.~Arleo, K.~J.~Eskola, H.~Paukkunen and C.~A.~Salgado,
%``Inclusive prompt photon production in nuclear collisions at RHIC and LHC,''
JHEP \textbf{04}, 055 (2011)

\bibitem{dEnterria:2012kvo} 
  D.~d'Enterria and J.~Rojo,
  %``Quantitative constraints on the gluon distribution function in the proton from collider isolated-photon data,''
  Nucl.\ Phys.\ B {\bf 860}, 311 (2012)


\bibitem{Helenius:2014qla}
I.~Helenius, K.~J.~Eskola and H.~Paukkunen,
%``Probing the small-$x$ nuclear gluon distributions with isolated photons at forward rapidities in p+Pb collisions at the LHC,''
JHEP \textbf{09}, 138 (2014)

%\cite{Klasen:2017dsy}
\bibitem{Klasen:2017dsy}
M.~Klasen, C.~Klein-B\"osing and H.~Poppenborg,
%``Prompt photon production and photon-jet correlations at the LHC,''
JHEP \textbf{03}, 081 (2018)

\bibitem{Goharipour:2018sip}
M.~Goharipour and S.~Rostami,
%``Probing nuclear modifications of parton distribution functions through the isolated prompt photon production at energies available at the CERN Large Hadron Collider,''
Phys. Rev. C \textbf{99}, no.5, 055206 (2019)

%%%%%%%%%%%%%%% Dipolos

\bibitem{Kopeliovich:1998nw}
B.~Z.~Kopeliovich, A.~V.~Tarasov and A.~Schafer,
%``Bremsstrahlung of a quark propagating through a nucleus,''
Phys. Rev. C \textbf{59}, 1609-1619 (1999)


\bibitem{Gelis:2002ki}   
  F.~Gelis and J.~Jalilian-Marian,
  %``Photon production in high-energy proton nucleus collisions,''
  Phys.\ Rev.\ D {\bf 66}, 014021 (2002).



\bibitem{Kopeliovich:2007yva}
B.~Z.~Kopeliovich, A.~H.~Rezaeian, H.~J.~Pirner and I.~Schmidt,
%``Direct photons and dileptons via color dipoles,''
Phys. Lett. B \textbf{653}, 210-215 (2007)



  
\bibitem{Kopeliovich:2009yw}
B.~Z.~Kopeliovich, E.~Levin, A.~H.~Rezaeian and I.~Schmidt,
%``Direct photons at forward rapidities in high-energy pp collisions,''
Phys. Lett. B \textbf{675}, 190-195 (2009)

\bibitem{Machado:2008nz}
M.~V.~T.~Machado and C.~B.~Mariotto,
%``Investigating the high energy QCD approaches for prompt photon production at the LHC,''
Eur. Phys. J. C \textbf{61}, 871-878 (2009)



\bibitem{amir} 
  J.~Jalilian-Marian and A.~H.~Rezaeian,
  %``Prompt photon production and photon-hadron correlations at RHIC and the LHC from the Color Glass Condensate,''
  Phys.\ Rev.\ D {\bf 86}, 034016 (2012); \\ 
  A.~H.~Rezaeian,
  %``Semi-inclusive photon-hadron production in pp and pA collisions at RHIC and LHC,''
  Phys.\ Rev.\ D {\bf 86}, 094016 (2012).

  

\bibitem{Basso:2015pba} 
  E.~Basso, V.~P.~Goncalves, J.~Nemchik, R.~Pasechnik and M.~Sumbera,
  %``Drell-Yan phenomenology in the color dipole picture revisited,''
  Phys.\ Rev.\ D {\bf 93}, no. 3, 034023 (2016)
  %doi:10.1103/PhysRevD.93.034023
  %[arXiv:1510.00650 [hep-ph]].



  
\bibitem{Ducloue:2017kkq} 
  B.~Ducloue, T.~Lappi and H.~Mantysaari,
  %``Isolated photon production in proton-nucleus collisions at forward rapidity,''
  Phys.\ Rev.\ D {\bf 97}, no. 5, 054023 (2018).


\bibitem{Benic:2016uku}
S.~Benic, K.~Fukushima, O.~Garcia-Montero and R.~Venugopalan,
%``Probing gluon saturation with next-to-leading order photon production at central rapidities in proton-nucleus collisions,''
JHEP \textbf{01}, 115 (2017)




\bibitem{Benic:2017znu}
S.~Beni\'c and A.~Dumitru,
%``Prompt photon - jet angular correlations at central rapidities in p+A collisions,''
Phys. Rev. D \textbf{97}, no.1, 014012 (2018)

\bibitem{Benic:2018hvb} 
  S.~Benić, K.~Fukushima, O.~Garcia-Montero and R.~Venugopalan,
  %``Constraining unintegrated gluon distributions from inclusive photon production in proton–proton collisions at the LHC,''
  Phys.\ Lett.\ B {\bf 791}, 11 (2019).
%  doi:10.1016/j.physletb.2019.02.007


\bibitem{Goncalves:2020tvh}
V.~P.~Goncalves, Y.~Lima, R.~Pasechnik and M.~\v{S}umbera,
%``Isolated photon production and pion-photon correlations in high-energy $pp$ and $pA$ collisions,''
Phys. Rev. D \textbf{101}, no.9, 094019 (2020)


\bibitem{SampaiodosSantos:2020lte}
G.~Sampaio dos Santos, G.~Gil da Silveira and M.~V.~T.~Machado,
%``The color dipole picture for prompt photon production in $pp$ and $pPb$ collisions at the CERN-LHC,''
Eur. Phys. J. C \textbf{80}, no.9, 812 (2020)


\bibitem{Golec-Biernat:2020cah}
K.~Golec-Biernat, L.~Motyka and T.~Stebel,
%``Prompt photon production in proton collisions as a probe of parton scattering in high energy limit,''
Phys. Rev. D \textbf{103}, no.3, 034013 (2021)


\bibitem{Benic:2022ixp}
S.~Beni\'c, O.~Garcia-Montero and A.~Perkov,
%``Isolated photon-hadron production in high energy pp and pA collisions at RHIC and LHC,''
Phys. Rev. D \textbf{105}, no.11, 114052 (2022)



  \bibitem{hdqcd} 
  F.~Gelis, E.~Iancu, J.~Jalilian-Marian and R.~Venugopalan,
    Ann.\ Rev.\ Nucl.\ Part.\ Sci.\  {\bf 60}, 463 (2010);
  H.~Weigert,  Prog.\ Part.\ Nucl.\ Phys.\  {\bf 55}, 461 (2005); J.~Jalilian-Marian and Y.~V.~Kovchegov, Prog.\ Part.\ Nucl.\ Phys.\  {\bf 56}, 104 (2006); A.~Morreale and F.~Salazar,
%``Mining for Gluon Saturation at Colliders,''
Universe \textbf{7}, no.8, 312 (2021).


\bibitem{CGC}  
J. Jalilian-Marian, A. Kovner, L. McLerran, and H. Weigert, Phys. Rev. D {\bf 55}, 5414 (1997);\\
J. Jalilian-Marian, A. Kovner, and  H.
Weigert, Phys. Rev. D {\bf 59}, 014014 (1999), {\it ibid.} {\bf 59}, 014015 (1999),
{\it ibid.} {\bf 59}  034007 (1999);\\
A. Kovner, J. Guilherme Milhano, and  H. Weigert, Phys. Rev. D {\bf 62}, 114005 (2000);\\
H. Weigert, Nucl. Phys. {\bf A703}, 823 (2002);\\
E. Iancu, A. Leonidov, and L. McLerran,
Nucl.Phys. {\bf A692}, 583 (2001);\\
E. Ferreiro, E. Iancu, A. Leonidov, and L. McLerran,
Nucl. Phys. {\bf A701}, 489 (2002).



\bibitem{BAL}  I. I. Balitsky,   Nucl. Phys. {\bf  B463}, 99 (1996); Phys. Rev. Lett. {\bf 81}, 2024 (1998); Phys. Rev. D  {\bf 60}, 014020 (1999); I. I. Balitsky,   Phys. Lett. B  {\bf 518}, 235 (2001);   I.I. Balitsky and  A.V. Belitsky, Nucl. Phys. {\bf B629}, 290  (2002). 
 

\bibitem{KOVCHEGOV}  
Y.V. Kovchegov,  Phys. Rev. D {\bf 60},  034008 (1999);  Phys. Rev. D {\bf 61} 074018 (2000). 



\bibitem{Ma:2018bax}
Y.~Q.~Ma, P.~Tribedy, R.~Venugopalan and K.~Watanabe,
%``Event engineering studies for heavy flavor production and hadronization in high multiplicity hadron-hadron and hadron-nucleus collisions,''
Phys. Rev. D \textbf{98}, no.7, 074025 (2018)





\bibitem{Levin:2019fvb}
E.~Levin, I.~Schmidt and M.~Siddikov,
%``Multiplicity dependence of quarkonia production in the CGC approach,''
Eur. Phys. J. C \textbf{80}, no.6, 560 (2020)


\bibitem{Kopeliovich:2019phc}
B.~Z.~Kopeliovich, H.~J.~Pirner, I.~K.~Potashnikova, K.~Reygers and I.~Schmidt,
%``Heavy quarkonium in the saturated environment of high-multiplicity $pp$ collisions,''
Phys. Rev. D \textbf{101}, no.5, 054023 (2020)


\bibitem{Gotsman:2020ubn}
E.~Gotsman and E.~Levin,
%``High energy QCD: multiplicity dependence of quarkonia production,''
Eur. Phys. J. C \textbf{81}, no.2, 99 (2021)



\bibitem{Siddikov:2020lnq}
M.~Siddikov and I.~Schmidt,
%``Multiplicity dependence of \ensuremath{\chi}c and \ensuremath{\chi}b meson production,''
Phys. Rev. D \textbf{104}, no.1, 016023 (2021)


\bibitem{Siddikov:2021cgd}
M.~Siddikov and I.~Schmidt,
%``Strangeness production in high-multiplicity events,''
Phys. Rev. D \textbf{104}, no.1, 016024 (2021)




\bibitem{Stebel:2021bbn}
T.~Stebel and K.~Watanabe,
%``J/\ensuremath{\psi} polarization in high multiplicity pp and pA collisions: CGC\,+\,NRQCD approach,''
Phys. Rev. D \textbf{104}, no.3, 034004 (2021)



 



\bibitem{Salazar:2021mpv}
F.~Salazar, B.~Schenke and A.~Soto-Ontoso,
%``Accessing subnuclear fluctuations and saturation with multiplicity dependent J/\ensuremath{\psi} production in p+p and p+Pb collisions,''
Phys. Lett. B \textbf{827}, 136952 (2022)


\bibitem{Lima:2022mol}
Y.~N.~Lima, A.~V.~Giannini and V.~P.~Goncalves,
%``Kaon production in high multiplicity events at the Large Hadron Collider,''
Phys. Rev. C \textbf{106}, no.6, 065206 (2022)







\bibitem{ALICE:2015ikl}
J.~Adam \textit{et al.} [ALICE],
%``Measurement of charm and beauty production at central rapidity versus charged-particle multiplicity in proton-proton collisions at $ \sqrt{s}=7 $ TeV,''
JHEP \textbf{09}, 148 (2015)


\bibitem{ALICE:2017wet}
D.~Adamov\'a \textit{et al.} [ALICE],
%``J/$\psi$ production as a function of charged-particle pseudorapidity density in p-Pb collisions at $\sqrt{s_{\rm NN}} = 5.02$ TeV,''
Phys. Lett. B \textbf{776}, 91-104 (2018)

\bibitem{STAR:2018smh}
J.~Adam \textit{et al.} [STAR],
%``$J/\psi$ production cross section and its dependence on charged-particle multiplicity in $p + p$ collisions at $\sqrt{s}$ = 200 GeV,''
Phys. Lett. B \textbf{786}, 87-93 (2018)


\bibitem{ALICE:2020msa}
S.~Acharya \textit{et al.} [ALICE],
%``Multiplicity dependence of J/$\psi$ production at midrapidity in pp collisions at $\sqrt{s}$ = 13 TeV,''
Phys. Lett. B \textbf{810}, 135758 (2020)

\bibitem{ALICE:2020eji}
S.~Acharya \textit{et al.} [ALICE],
%``J/$\psi$ production as a function of charged-particle multiplicity in p-Pb collisions at $\sqrt{\textit{s}_{\rm NN}}~=~8.16$ TeV,''
JHEP \textbf{09}, 162 (2020)





\bibitem{ALICECollaboration:2020}
ALICE Collaboration,
%\emph{Multiplicity dependence of (multi-)strange hadron production in proton-proton collisions at $\sqrt{s} = 13$ TeV},
%\href{https://doi.org/10.1140/epjc/s10052-020-7673-8}
Eur. Phys. J. C \textbf{80}, 167 (2020).


\bibitem{ALICE:2021zkd}
S.~Acharya \textit{et al.} [ALICE],
%``Forward rapidity J/\ensuremath{\psi} production as a function of charged-particle multiplicity in pp collisions at $ \sqrt{s} $ = 5.02 and 13 TeV,''
JHEP \textbf{06}, 015 (2022)





%%%%%%%%%%%%%%%%%%%%%%%%%%   NLO %%%%%%%%%%%%%%%%%%%%%%%%%%%%%%%%%%%%%%%

\bibitem{Chirilli:2011km}
G.~A.~Chirilli, B.~W.~Xiao and F.~Yuan,
%``One-loop Factorization for Inclusive Hadron Production in $pA$ Collisions in the Saturation Formalism,''
Phys. Rev. Lett. \textbf{108}, 122301 (2012)

\bibitem{Balitsky:2012bs}
I.~Balitsky and G.~A.~Chirilli,
%``Photon impact factor and $k_T$-factorization for DIS in the next-to-leading order,''
Phys. Rev. D \textbf{87}, no.1, 014013 (2013)

\bibitem{Altinoluk:2014eka}
T.~Altinoluk, N.~Armesto, G.~Beuf, A.~Kovner and M.~Lublinsky,
%``Single-inclusive particle production in proton-nucleus collisions at next-to-leading order in the hybrid formalism,''
Phys. Rev. D \textbf{91}, no.9, 094016 (2015)



\bibitem{Beuf:2016wdz}
G.~Beuf,
%``Dipole factorization for DIS at NLO: Loop correction to the $\gamma^*_{T,L}\to q\overline q$ light-front wave functions,''
Phys. Rev. D \textbf{94}, no.5, 054016 (2016)


\bibitem{Iancu:2016vyg}
E.~Iancu, A.~H.~Mueller and D.~N.~Triantafyllopoulos,
%``CGC factorization for forward particle production in proton-nucleus collisions at next-to-leading order,''
JHEP \textbf{12}, 041 (2016)



\bibitem{Lappi:2016fmu}
T.~Lappi and H.~M\"antysaari,
%``Next-to-leading order Balitsky-Kovchegov equation with resummation,''
Phys. Rev. D \textbf{93}, no.9, 094004 (2016)

\bibitem{Boussarie:2016ogo}
R.~Boussarie, A.~V.~Grabovsky, L.~Szymanowski and S.~Wallon,
%``On the one loop $ {\gamma}^{\left(\ast \right)}\to q\overline{q} $ impact factor and the exclusive diffractive cross sections for the production of two or three jets,''
JHEP \textbf{11}, 149 (2016)

\bibitem{Beuf:2017bpd}
G.~Beuf,
%``Dipole factorization for DIS at NLO: Combining the $q\bar{q}$ and $q\bar{q}g$ contributions,''
Phys. Rev. D \textbf{96}, no.7, 074033 (2017)

\bibitem{Boussarie:2016bkq}
R.~Boussarie, A.~V.~Grabovsky, D.~Y.~Ivanov, L.~Szymanowski and S.~Wallon,
%``Next-to-Leading Order Computation of Exclusive Diffractive Light Vector Meson Production in a Saturation Framework,''
Phys. Rev. Lett. \textbf{119}, no.7, 072002 (2017)


\bibitem{Caucal:2021ent}
P.~Caucal, F.~Salazar and R.~Venugopalan,
%``Dijet impact factor in DIS at next-to-leading order in the Color Glass Condensate,''
JHEP \textbf{11}, 222 (2021)

\bibitem{Liu:2022ijp}
H.~y.~Liu, K.~Xie, Z.~Kang and X.~Liu,
%``Single inclusive jet production in pA collisions at NLO in the small-x regime,''
JHEP \textbf{07}, 041 (2022)




\bibitem{Taels:2022tza}
P.~Taels, T.~Altinoluk, G.~Beuf and C.~Marquet,
%``Dijet photoproduction at low x at next-to-leading order and its back-to-back limit,''
JHEP \textbf{10}, 184 (2022)

\bibitem{Bergabo:2022tcu}
F.~Bergabo and J.~Jalilian-Marian,
%``One-loop corrections to dihadron production in DIS at small x,''
Phys. Rev. D \textbf{106}, no.5, 054035 (2022)

\bibitem{Fucilla:2022wcg}
M.~Fucilla, A.~V.~Grabovsky, E.~Li, L.~Szymanowski and S.~Wallon,
%``NLO computation of diffractive di-hadron production in a saturation framework,''
JHEP \textbf{03}, 159 (2023)

\bibitem{Bergabo:2022zhe}
F.~Bergabo and J.~Jalilian-Marian,
%``Single inclusive hadron production in DIS at small x: next to leading order corrections,''
JHEP \textbf{01}, 095 (2023)




\bibitem{Bergabo:2023wed}
F.~Bergabo and J.~Jalilian-Marian,
%``Dihadron production in DIS at small x at next-to-leading order: Transverse photons,''
Phys. Rev. D \textbf{107}, no.5, 054036 (2023)

\bibitem{Caucal:2023nci}
P.~Caucal, F.~Salazar, B.~Schenke, T.~Stebel and R.~Venugopalan,
%``Back-to-back inclusive dijets in DIS at small x: gluon Weizs\"acker-Williams distribution at NLO,''
JHEP \textbf{08}, 062 (2023)

\bibitem{Altinoluk:2023hfz}
T.~Altinoluk, N.~Armesto, A.~Kovner and M.~Lublinsky,
%``Single inclusive particle production at next-to-leading order in proton-nucleus collisions at forward rapidities: Hybrid approach meets TMD factorization,''
Phys. Rev. D \textbf{108}, no.7, 074003 (2023)


\bibitem{Fucilla:2023mkl}
M.~Fucilla, A.~Grabovsky, E.~Li, L.~Szymanowski and S.~Wallon,
%``Diffractive single hadron production in a saturation framework at the NLO,''
[arXiv:2310.11066 [hep-ph]].

\bibitem{Taels:2023czt}
P.~Taels,
%``Forward production of a Drell-Yan pair and a jet at small x at next-to-leading order,''
JHEP \textbf{01}, 005 (2024)

\bibitem{cteq} 
  H.~L.~Lai, M.~Guzzi, J.~Huston, Z.~Li, P.~M.~Nadolsky, J.~Pumplin and C.-P.~Yuan,
  %``New parton distributions for collider physics,''
  Phys.\ Rev.\ D {\bf 82}, 074024 (2010).



\bibitem{Albacete:2007yr}
J.~L.~Albacete and Y.~V.~Kovchegov,
%``Solving high energy evolution equation including running coupling corrections,''
Phys. Rev. D \textbf{75}, 125021 (2007)

\bibitem{Albacete:2009fh}
J.~L.~Albacete, N.~Armesto, J.~G.~Milhano and C.~A.~Salgado,
%``Non-linear QCD meets data: A Global analysis of lepton-proton scattering with running coupling BK evolution,''
Phys. Rev. D \textbf{80}, 034031 (2009)



\bibitem{Albacete:2010sy}
J.~L.~Albacete, N.~Armesto, J.~G.~Milhano, P.~Quiroga-Arias and C.~A.~Salgado,
%``AAMQS: A non-linear QCD analysis of new HERA data at small-x including heavy quarks,''
Eur. Phys. J. C \textbf{71}, 1705 (2011)






\bibitem{Albacete:2012xq}
J.~L.~Albacete, A.~Dumitru, H.~Fujii and Y.~Nara,
%``CGC predictions for p + Pb collisions at the LHC,''
Nucl. Phys. A \textbf{897}, 1-27 (2013)



\bibitem{ATLAS:2016}
ATLAS Collaboration,
%``Measurement of the inclusive isolated prompt photon cross section in pp collisions at \sqrt{s}=8 TeV with the ATLAS detector''
JHEP \textbf{08}, 005 (2016)



\bibitem{Lappi:2011gu}
T.~Lappi,
%``Energy dependence of the saturation scale and the charged multiplicity in pp and AA collisions,''
Eur. Phys. J. C \textbf{71}, 1699 (2011)



\end{thebibliography}
\end{document}